\documentclass[11pt]{article}
\usepackage[a4paper,margin=1.0in,footskip=0.5in]{geometry}
\usepackage{authblk}
\usepackage{graphics,graphicx,epsfig}
\usepackage{multirow}
\usepackage{breqn}
\usepackage[round]{natbib}
\usepackage{amssymb,amsmath,amsfonts,eurosym,geometry,ulem,graphicx,caption,color,setspace,sectsty,comment,footmisc,caption,natbib,pdflscape,subfigure,array,hyperref}

\newcolumntype{L}[1]{>{\raggedright\let\newline\\arraybackslash\hspace{0pt}}m{#1}}
\newcolumntype{C}[1]{>{\centering\let\newline\\arraybackslash\hspace{0pt}}m{#1}}
\newcolumntype{R}[1]{>{\raggedleft\let\newline\\arraybackslash\hspace{0pt}}m{#1}}

\title{Does Reliable Electricity Mean Lesser Agricultural Labor Wages? Evidence from Indian Villages}

\author[a]{Suryadeepto Nag}
\affil[a]{Department of Humanities and Social Sciences, Indian Institute of Science Education and Research Pune, Pashan, Pune - 411008, India. E-mail: nag.suryadeepto@gmail.com}

\begin{document}
\maketitle
\begin{abstract}
    Using a panel of 1,171 villages in rural India that were surveyed in the India Human Development Surveys, I perform a difference-in-differences analysis to find that improvements in electricity reliability have a negative effect on the increase in casual agricultural labor wage rates. Changes in men's wage rates are found to be affected more adversely than women's, resulting in a smaller widening of the gender wage gap. I find that better electricity reliability reduces the time spent by women in fuel collection substantially which could potentially increase labor supply. The demand for labor remains unaffected by reliability, which could lead the surplus in labor supply to cause wage rates to stunt. However, I show that electrical appliances such as groundwater pumps considerably increase labor demand indicating that governments could target increasing the adoption of electric pumps along with bettering the quality of electricity to absorb the surplus labor into agriculture.  
\end{abstract}



\textbf{Acknowledgements}: I thank David I. Stern for several helpful discussions and detailed comments.

\setcounter{page}{0}
\thispagestyle{empty}

\pagebreak \newpage

\doublespacing

\section{Introduction} \label{sec:introduction}
The question of whether advancements in electrification have effects on agriculture has been a longstanding one \citep{barnes1986impact}. In the late twentieth century, several countries promoted developments in electricity infrastructure to increase the adoption of electrical farm equipment and agricultural productivity. However, recent studies \citep{khandker2013welfare,kumar2018impact,nag2023are} have cast doubt on any electricity-induced benefits for agricultural income. Does this imply electricity access has no consequence on agriculture, or is it simply not significantly bettering productivity and income? In this paper, I study a less-studied dimension of electricity and agriculture -- the effect of electricity quality on agricultural labor wages.

\begin{table}[h]
\color{black}
    \centering
   \resizebox{\textwidth}{!}{ \begin{tabular}{cccccccccc}
    \hline 
     &&&&&&&&&\\
         && \multicolumn{8}{c}{Means (standard deviation)  }   \\ 
          &&&&&&&&&\\
          \cline{3-10}
      &&&&&&&&&\\
         && \multicolumn{2}{c}{2004-2005} && \multicolumn{2}{c}{2011-2012}&& \multicolumn{2}{c}{$\Delta$}\\
           &&&&&&&&&\\
         \cline{3-4} \cline{6-7} \cline{9-10}
 &&&&&&&&&\\
         && Increase in & Non-increase in && Increase in & Non-increase in &&Increase in & Non-increase in \\ 
         && Reliability & Reliability && Reliability & Reliability && Reliability & Reliability \\
          &&&&&&&&&\\
          \hline 
           &&&&&&&&&\\
         
         \multirow{1}{*}{Women's daily wage rate (2012 Rs.)} &  & 86.71  & 85.98  && 133.29  &  138.93  && 40.44  &47.31  \\
       && (40.2) & (41.72) && (57.2) & (69.97)&&  (46.99)   & (49.97)\\
         \multirow{1}{*}{Men's daily wage rate (2012 Rs.) }&  & 116.5  & 119.55  && 167.2  &  183.32  && 50.52  & 62.81 \\
       && (51.45) & (52.21) && (66.76) & (83.07)&&  (56.14)   &(61.62)\\
         
          \multirow{1}{*}{Average daily wage rate (2012 Rs.)}& & 104.93   & 105.42   && 150.5  &  161.63  && 44.8   & 55.45  \\
     && (47.08) & (46.46) && (59.37) & (73.87)&& (50.16)    &(54.64)\\
        &&&&&&&&&\\
         \hline 
    \end{tabular}}
    \caption{ \color{black}Casual agricultural labor wage rate statistics for villages where reliability improves and villages where it does not. Reliability is measured as the average number of hours of electricity available per day. Includes 1281 Indian villages. The wages have been standardized to the 2012 Rupee by multiplying the 2005 rates by the national consumer price index (reported by the World Bank). Averages are unweighted. The data are unweighted. Source: IHDS I, and IHDS II surveys. \color{black}}
    \label{tab:wages_stats}
    \color{black}
\end{table}
Taking a cursory glance at village-level data from the India Human Development Surveys (IHDS) \citep{IHDSI, IHDSII}, we can observe a strange phenomenon (Table \ref{tab:wages_stats}). Villages where the quality of electricity improved from 2004-5 to 2011-12, have lower agricultural labor wage rates in 2012 than villages where the quality of electricity worsened. Both types of villages have similar average wage rates in 2004-5, and there is a general trend of increasing wages. However, the increase is less in villages that observe an improvement in the hours of access, when compared to villages that don't (most of which observe a decline). The disparity is consistent in both men's and women's wages, although the difference is larger in men's wages. After controlling for inflation, I find that the increase in wage rates is close to 24\% higher in villages where the reliability worsened or stayed at the same level than in villages where the reliability improved. Although the observation may merely be a correlation, a causal link between electricity and agriculture, wherein better quality electricity drives down agricultural labor wages, may have implications for our understanding of the role of electricity interventions in agrarian economies.

 For my study, I use data from two panels of Indian villages surveyed as part of the India Human Development Surveys (IHDS), covering a period from 2004-5 to 2011-12. Among lower and lower-middle-income countries, India has been among the forerunners in the expansion of energy access, and studying the causal effects of reliability in India has consequences for several countries of the Global South expected to ramp up their power generation and distribution to meet electrification targets in the coming years. To investigate if improved electricity access is causally driving down wages, I employ a difference-in-differences design, while controlling for several village-level variables, to suitably identify the effects of reliability on the wage rates, and eliminate biases due to village- and time-fixed effects. My analysis is lent further aid by the rich IHDS survey data sets,  which have data on several variables both at the level of households and villages. The data, representative at the national level, includes several essential variables, apart from reliability, such as the time since when a village has electricity, and the fraction of households in the village that have access to electricity. The surveys also have data on various other characteristics such as the status of infrastructure in the village, proximity to banks and markets, the number of schools in the village, etc., making it easier to control for a large number of confounding variables.

The period of study (2004-5 to 2011-12) coincides with India's last-mile electrification efforts, with over 90\% villages already having been electrified prior to this period, and over 99\% by the end of it. Since most villages had already been connected to the grid, rural India provides a good stage to study the more intricate dimensions of electricity access such as availability and disruptions. According to data from the World Bank, in 2005, India constituted the largest rural population of the World with villages making up 71\% of the national population. At this time, agriculture, forestry, and fisheries made up over 15\% of the Indian GDP (over 100 billion 2021 USD in size), compared to the global average of 3.2\% share of the World GDP, making India an ideal site for investigating changes in agricultural practices and wages. 

I find that improvements in electricity quality result in a lower increase in casual agricultural labor wage rates. The effect is especially pronounced in men's wage rates, which consequently results in a smaller widening of the gender wage gap in these villages. On analyzing the time spent in fuel collection which could be a possible labor supply channel, I find that reliable electricity reduces the time burden of biomass collection, which could potentially increase labor supply. On the other hand, analyzing labor demand, I find that the demand for labor does not change with reliability. Since disguised unemployment and saturated labor demand are existing concerns in the Indian agricultural sector, households may not be able to reap reliability-linked benefits to income that are found elsewhere \cite{dang2019does, pepino2021does}. Instead, a possible increase in labor market participation could potentially hurt wage rates. However, I find that electrical farm machinery such as groundwater pumps have a large positive effect on the demand for labor. Therefore, policymakers could focus on greater investments in electric pumps or other alternate avenues that could increase labor demand and absorb the surplus labor, to help reduce what may be an electricity-induced aggravation of disguised unemployment.

The paper is organized as follows: In the next section, I discuss some theoretical arguments and the state of the literature. In the third section, I discuss the institutional context and present the data. This is followed by the empirical strategy in the fourth section, and the results in the fifth. In the final section, I discuss the conclusions.

\section{Theoretical Arguments and Evidence from the Literature} \label{sec:Theory_Literature}
Although several studies have looked at the effect of electricity quality and outages on industrial performance \citep{allcott2016electricity, maruyama2019underutilized, fakih2020effects}, few studies have investigated the economic benefits of reliability on well-being, in general \citep{nduhuura2021impacts, hussain2023understanding}, and lesser still on agriculture. Studies find mostly positive impacts of electricity quality on income including \citet{chakravorty2014does} in India, \citet{dang2019does} in Vietnam, and \citet{pepino2021does} in the Philippines. \citet{samad2016benefits}, however, find that electricity reliability reduces non-farm income in India (2\% reduction for every additional hour of power).\footnote{ \citet{samad2016benefits} are primarily interested in studying the effects of electricity access, and reliability is used as more of a control variable. The authors use propensity-score-weighted-regressions to remove selection bias in which households receive access, and the estimates for the effect of reliability may be biased by the weights. They also include households not connected to the grid as a control, assuming zero reliability, but these households may not benefit the same way as households who observe no changes in their existing reliability.}  Given that the main source of  non-farm income in rural India is agricultural labor, this would be consistent with a reliability-induced fall in wage rates that we see in Table \ref{tab:wages_stats}. 

There are several ways through which the quality of electricity could impact agricultural practices and production \citep{costantini2004social}, which could have an impact on the demand for labor. Studies have shown that electricity access increases the adoption of electric pumps \citep{barnes1986impact, smith2016rural}. If irrigation pumps are operated at fixed times in the day, irregularity of power could affect the usability of pumps, and agricultural households may be forced to employ conventional methods of manual irrigation or other machinery such as diesel pumps \citep{smith2016rural}. Agricultural mechanization may supplant human labor used in farms \citep{baur2023replacing}, and the replacement could reduce labor demand. However, the presence of the machines may otherwise also lead to an increased demand for workers to operate the machines or simply as a consequence of increased productivity or farming during dry seasons. In the Indian context, the latter is more likely. Reliability may also affect agriculture indirectly, by affecting decision-makers and laborers, which may alter the demand and supply of agricultural labor in villages. Better quality electricity could help households diversify their income through entrepreneurial options, as well as new employment opportunities that may have opened up because of better reliability. The demand for agricultural labor demand may also be affected by enhanced affluence, particularly from non-farm income increases, such as the effects observed by \citet{chakravorty2014does}. This may reduce households' reliance on agriculture, which may, in turn, reduce the demand for agricultural labor. 

In the context of labor supply, electricity access is also thought to have a positive impact on labor market participation \citep{dinkelman2011effects, salmon2016rural, rathi2018rural}. Labor supply could increase via a reduced time burden of domestic chores, especially for women. Since reliable electricity is essential for disruption-free electric lighting and refrigeration, better quality electricity could relieve individuals of the time spent in fuel collection \citep{samad2016benefits,njenga2021women}, and may enable them to participate in the labor market, which could increase the supply of labor.  Whether there is a surplus of agricultural labor and disguised unemployment \citep{robinson1936disguised} in India and other developing countries has been a topic of interest \citep{wellisz1968dual}. \cite{foster2010there} suggest that surplus labor already exists in Indian agriculture. Thus additional electricity-induced participation in casual agricultural labor markets is only likely to reduce the village-level wage rates if there isn't a proportional increase in the demand for labor. However, a reduction in the time burden of domestic chores may not necessarily encourage household members to engage in paid employment. In the case of agricultural households, this may result in members of the household supplying more labor to household farms, which could reduce the demand for hired labor.

The evidence on the relationship between electrification and agricultural labor has been mixed, although most studies look at the presence of electricity connections rather than the quality of electricity. In the context of casual labor, \citet{van2017long} study the impact of electricity access on labor supply and wage rates in India using a long period from 1982-1999. The authors find a negative impact on the number of days of casual wage work supplied by men, from both household and village-level electrification, although they do not find such effects for women. However, using data from harvest wage rates, the authors do not find a statistically significant impact of electrification on wage rates, either for men or women. To study village-level electrification, though, \citet{van2017long} use only one dimension of electricity -- the time since the village was first connected. In contrast, \citet{emran2018agricultural} use the fraction of connected households as a control variable in their study of the effects of agricultural productivity on hired labor and find a statistically significant negative effect associated with the fraction of households connected, implying that villages with better access to electricity saw agricultural households hiring less labor. Using solitary variables, as in the cases above, may lead to omitted variable biases and paint an incomplete picture. Therefore in my analysis, I complement the reliability variable with the fraction of households connected and also use the time since the village was first connected as a control to test for endogeneity in treatment assignment, to appropriately arrive at unbiased estimates of the effect of reliability.

\section{Setting and Data} \label{sec:setting_data} 
\subsection{Historical Context} \label{ss:history}
At the time of India's independence from the British Empire, in 1947, a majority of India's population was neither educated nor literate, with its rural population particularly worse off. In the absence of industries, rural India was overwhelmingly agrarian. In 1951, agriculture and allied sectors contributed to 51.9\% of India's GDP \citep{wagh2016agricultural}. While that number has since fallen sharply to 17\% in 2014-15, a majority of Indians continue to remain employed in agriculture. Therefore, it comes as little surprise that Government efforts in the second half of the 20th century toward electrifying villages were focused on connecting farms, rather than households. 

\begin{figure}[h]
\centering
        \includegraphics[width=0.7\linewidth]{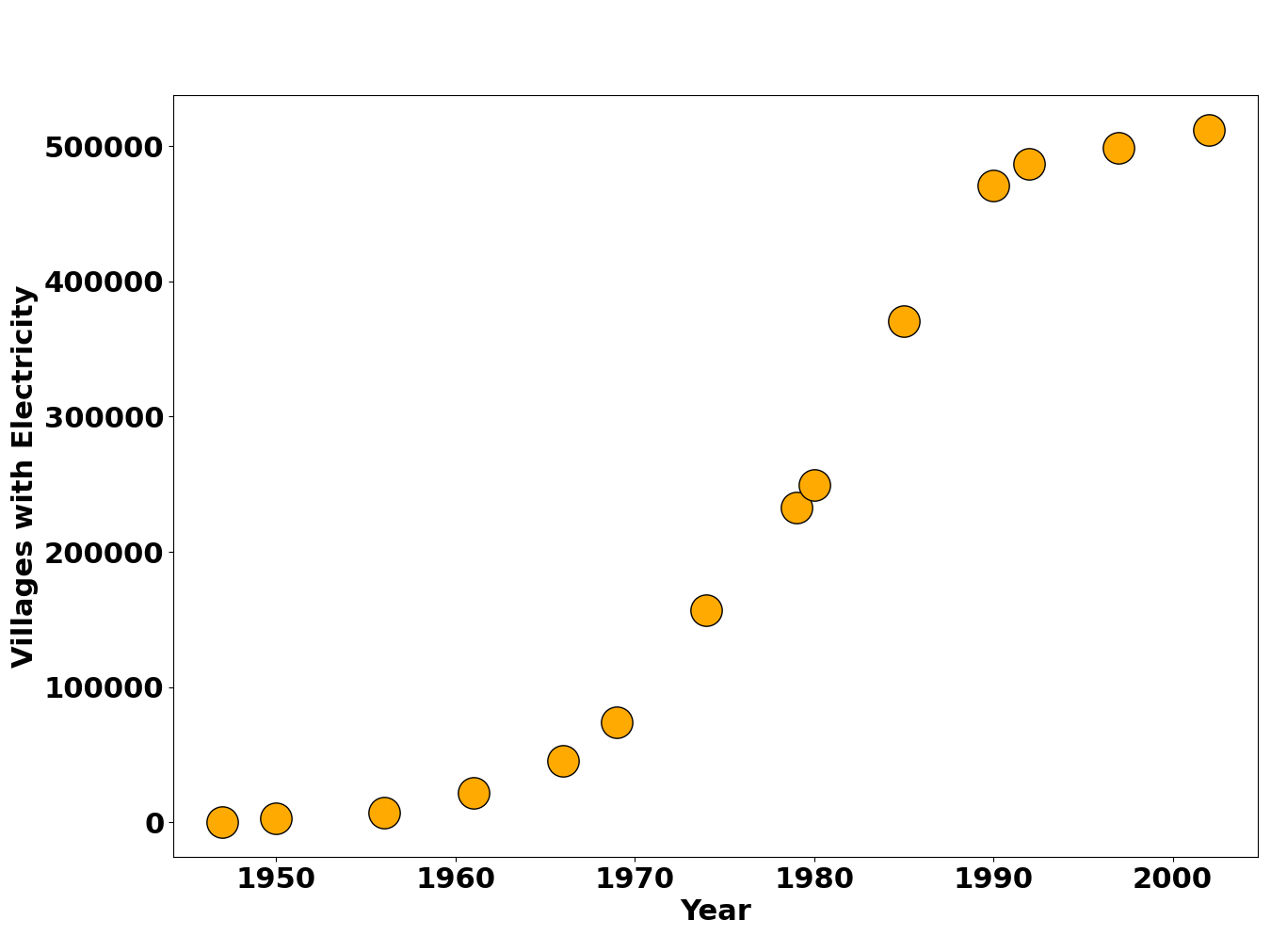}
        \caption{Time series of the total number of Indian villages with electricity connections in India since 1947. Source: \cite{growthbook2021}}
        \label{subfig:vill_con_yrs}
\end{figure}
\begin{figure}
    \centering
    \includegraphics[width=0.7\linewidth]{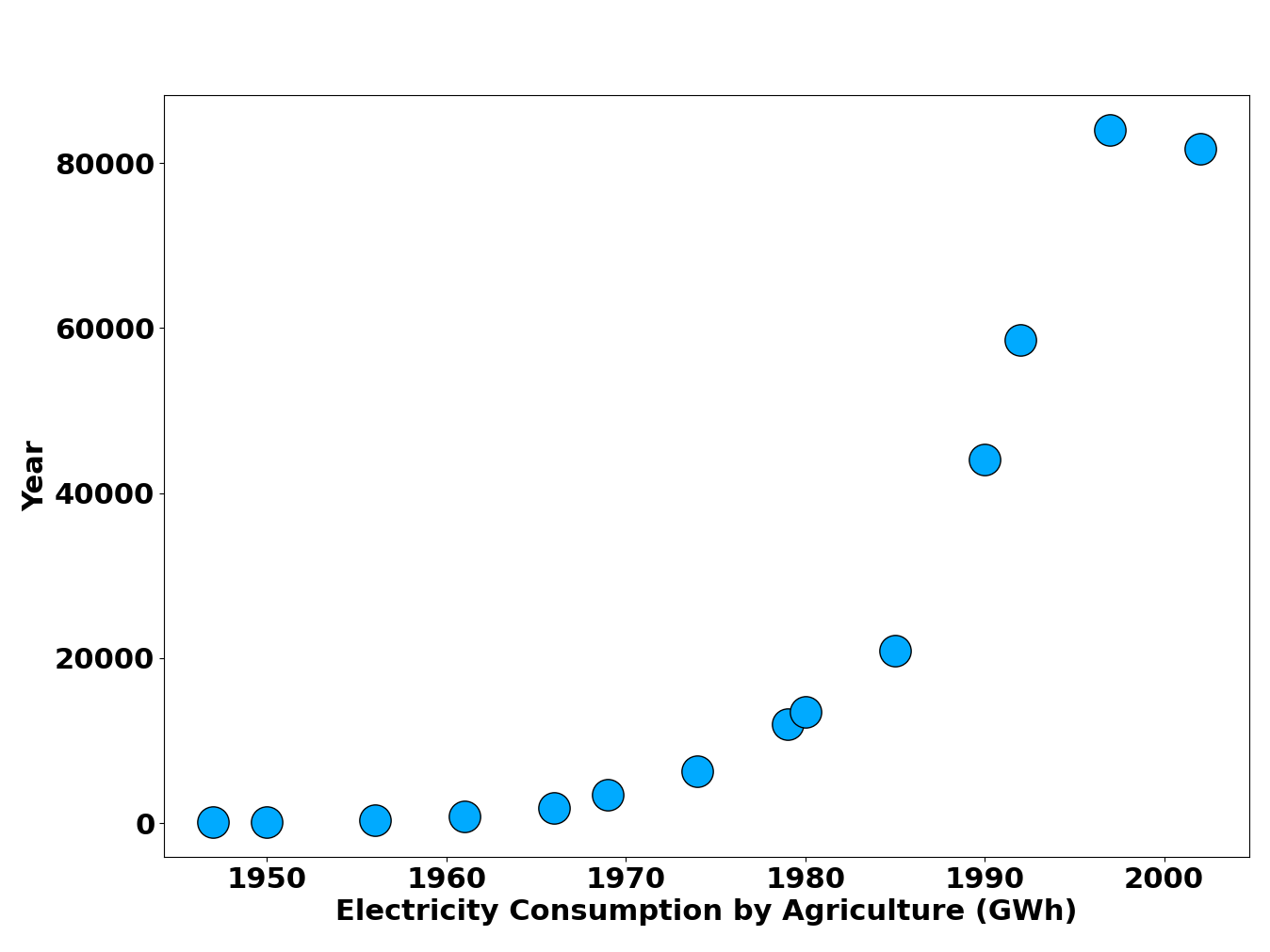}
    \caption{Time series of the total electric power consumption by the agricultural sector in India since 1947. Source: \cite{growthbook2021}}
    \label{subfig:ag_con_yrs}
\end{figure}

Investments in bringing electric pumps to villages were further accelerated by the famine in Bihar in 1966-1976, and the drought in Maharashtra in 1970-1973 \citep{dubhashi1992drought}. In 1969, the Ministry of Power established the Rural Electrification Corporation to oversee village electrification, with the specific purpose of aiding State Electricity Boards in facilitating the adoption of electric groundwater pumps in villages to increase agricultural productivity.  The droughts and famines of the 1960s and 1970s, ultimately along with the Green Revolution in the 1970s, led to an expeditious increase in the rate of village electrification in India, as can be seen in Figure \ref{subfig:vill_con_yrs}. Despite the growth in the number of villages connected, power consumption by the agricultural sector remained low, and consumption only crossed 10,000GWh in the late 1970s (Figure \ref{subfig:ag_con_yrs}), with an annual consumption of 12,028GWh in 1979, at the end of the Fifth 5-year Plan.\footnote{Reported progress at the end of the Indian financial year (31$^{st}$ March) on the final year of the Plan.} By the end of the Seventh Plan, in 1997, electricity consumption had grown sevenfold. The growth in agricultural power consumption and the overall installed capacity \citep{growthbook2021} may imply that even though many villages were connected to the grid in the mid-late 1900s, the quality and quantity may have been subpar, or it may mean that it takes a longer time before households can make use of their electricity connections \citep{nag2023are}.

Village electrification continued at a steady rate till the turn of the century, and by 2002, approximately 90\% villages\footnote{According to the 2001 census, India had 593,732 inhabited villages in 2001.} had been connected \citep{growthbook2021}, although a large number of households in these villages were yet to be connected as of 2004-5 \citep{IHDSI}. In fact, until the year 2003, power consumption by the agricultural sector exceeded domestic power consumption at the national level \citep{growthbook2021}. It is only in 2005 that the government began prioritizing household connections, and launched the Rajiv Gandhi Grameen Vidyutikaran Yojana (RGGVY) to connect the remaining unelectrified households. It is during this period that my study is set.

\subsection{Study Period and Dataset} \label{ss:data}
For my analysis, I use data from two waves of the India Human Development Survey \citep{IHDSI, IHDSII}. The first round of the survey was conducted in 2004-2005, and the second round was conducted in 2011-2012. This makes for a suitable period, as most villages had already been connected, allowing for a larger sample of villages where changes in reliability can be observed and studied. However, since a number of households would be connected in this period, my analysis would need to control for the fraction of households in a village that were connected to the grid. According to the \citet{growthbook2021}, transmission and distribution losses were considerably higher than average in this period (31.25\% in 2004-5 which reduced to 23.65\% by 2011-12). This would have an impact on the reliability of electricity and induce variation in the sample. This period is also roughly 15 years after the shock of economic liberalization in India, and the benefits to economic growth had already set in, distinct from earlier work that investigated wage rates before and after the economic liberalization of 1991 \citep{van2017long}.

The IHDS-I survey covered 41,554 households, and IHDS-II covered 42,152 households in total from 384 districts, 1,503 villages, and 971 urban blocks. Since I am primarily interested in village-level quantities like labor wage rates, I restrict my preliminary analysis to the village questionnaires and data. In all, I construct a panel of 1406 villages from the two survey rounds. I further use household-level data to study labor demand and supply in an attempt to explain the phenomena observed in the village-level analysis. Although there are differences in some of the questions asked in each survey, most of the variables of interest are present in both rounds.

The IHDS surveys make for an excellent data set to study the impact of electrical reliability on agricultural labor wages. The survey has detailed data on several insightful dimensions of village-level electricity. The survey tells me whether a village has electricity, what fraction of households in the village have electricity, when the village was first connected to the grid, and the average number of hours in a day that a village receives power. Similarly, the surveys also have data on the casual agricultural wage rates for sowing, and harvesting, for both men and women and for both major cropping seasons in India - Kharif (summer/monsoon) and Rabi(winter). Since the surveys also have household-level data, I can further investigate the labor demand and supply channels contributing to the trends observed in agricultural wages, at a more microscopic level, by studying changes in agricultural households.

\subsection{Descriptive Statistics} \label{ss:desc_stats}
Since I work with observed data, rather than an experiment, whether a village experienced improvements or declines in reliability need not necessarily have been determined by randomization. Therefore, it is important to check if the assignment of villages to different treatment doses i.e., different levels of improvement or declines in electricity reliability were influenced by other factors. The ideal way to check whether the treatment is ``As good as random" is to see if there are parallel trends in the treatment groups in the pre-treatment periods. However, in the absence of multiple pre-treatment periods, the best I can do is make the observation that the pre-treatment levels of wages are essentially indistinguishable in the pre-treatment period (Table \ref{tab:wages_stats}).

\begin{table}[h]
    \centering
   \resizebox{\textwidth}{!}{ \begin{tabular}{cccccccc}
    \hline 
    &&&&&&&\\
         & \multicolumn{5}{c}{Means (standard deviation)  }&&   \\ &&&&&&&\\\cline{2-6}
         &&&&&&&\\
         & \multicolumn{2}{c}{2004-2005} && \multicolumn{2}{c}{2011-2012}&& \\ 
         &&&&&&&Correlation of\\\cline{2-3} \cline{5-6} 
         &&&&&&&$\Delta$Reliability with\\
         & Positive Treatment & Negative Treatment && Positive Treatment & Negative Treatment &&Initial levels (2004-5) \\ 
         &&&&&&&\\\hline 
         &&&&&&&\\
         Percentage of households with Electricity access (\%) & 69.98 & 71.50 && 80.36 &  79.90 && 0.00\\
         & (32.83) & (30.19) && (25.99) & (24.86)&&\\
         Presence of Metalled Roads$^\dagger$ & 0.62 & 0.73 && 0.86 & 0.90&&-0.09$^{**}$\\
         & (0.49) & (0.45) && (0.35) & (0.31) &&\\
        Distance to the nearest bank branch/credit cooperative (km) & 5.27 & 4.03 && 5.44 & 4.54 &&0.10$^{***}$\\ 
         & (6.05) & (4.61) && (5.99) & (4.95) &&\\
         Distance to the closest market (km) & 2.37  &  2.38   &&  2.43  &  2.78  && 0.00\\
         & (4.69) & (4.33) && (5.26) & (5.50) &&\\
        Presence of NGO/Development Organization$^\dagger$ &  0.11     &  0.15      && 0.14     &   0.14    && -0.07$^{*}$ \\
         &  (0.31)    &  (0.36)     && (0.35)    &  (0.35)  && \\
         Presence of Primary Healthcare Center$^\dagger$   & 0.11 & 0.17 && 0.12  & 0.15 &&-0.09$^{**}$\\
          & (0.32) & (0.44) && (0.33) & (0.36)&&\\
           Number of Government Primary Schools$^\dagger$ & 1.54 & 1.95&& 1.67  & 1.52 &&-0.17$^{***}$\\
         & (1.34) & (1.89) && (1.54) & (1.78) &&\\
        Number of Government Middle Schools$^\dagger$ & 0.70 & 0.74&& 0.88  & 0.89 &&-0.09$^{**}$\\
         & (0.68) & (0.74) && (0.73) & (0.72) &&\\
         Number of Government Secondary Schools$^\dagger$ &0.28& 0.34&& 0.38  & 0.42 &&-0.05\\
         & (0.51) & (0.53) && (0.66) & (0.61) &&\\
               Number of Government Higher Secondary Schools$^\dagger$ & 0.11 & 0.14&& 0.18  & 0.17 &&-0.03\\
         & (0.35) & (0.37) && (0.44) & (0.43) &&\\
         &&&&&&&\\
         \hline 
    \end{tabular}}
    \caption{Descriptive statistics for treatment and control Villages 2004-2005 and 2011-2012. ``Positive Treatment" villages refer to those villages which see an improvement in the average number of hours of electricity received in a day, and ``Negative Treatment" villages refer to those which do not (These include control villages where there is no change). Correlation refers to Pearson's correlation coefficient, with the two-sided alternate hypothesis. Includes 1254 villages. The data are unweighted. Source: IHDS I, and IHDS II surveys.\\
    \small{$^\dagger$ Dummy variable which takes 1 for ``yes" and 0 for ``no".}\\ \small{*Significant at the 5\% level, 
    **Significant at the 1\% level, 
    ***Significant at the 0.1\% level}}
    \label{tab:hh_rel}
\end{table}
\begin{figure}[h]
    \centering
        \includegraphics[width=0.7\linewidth]{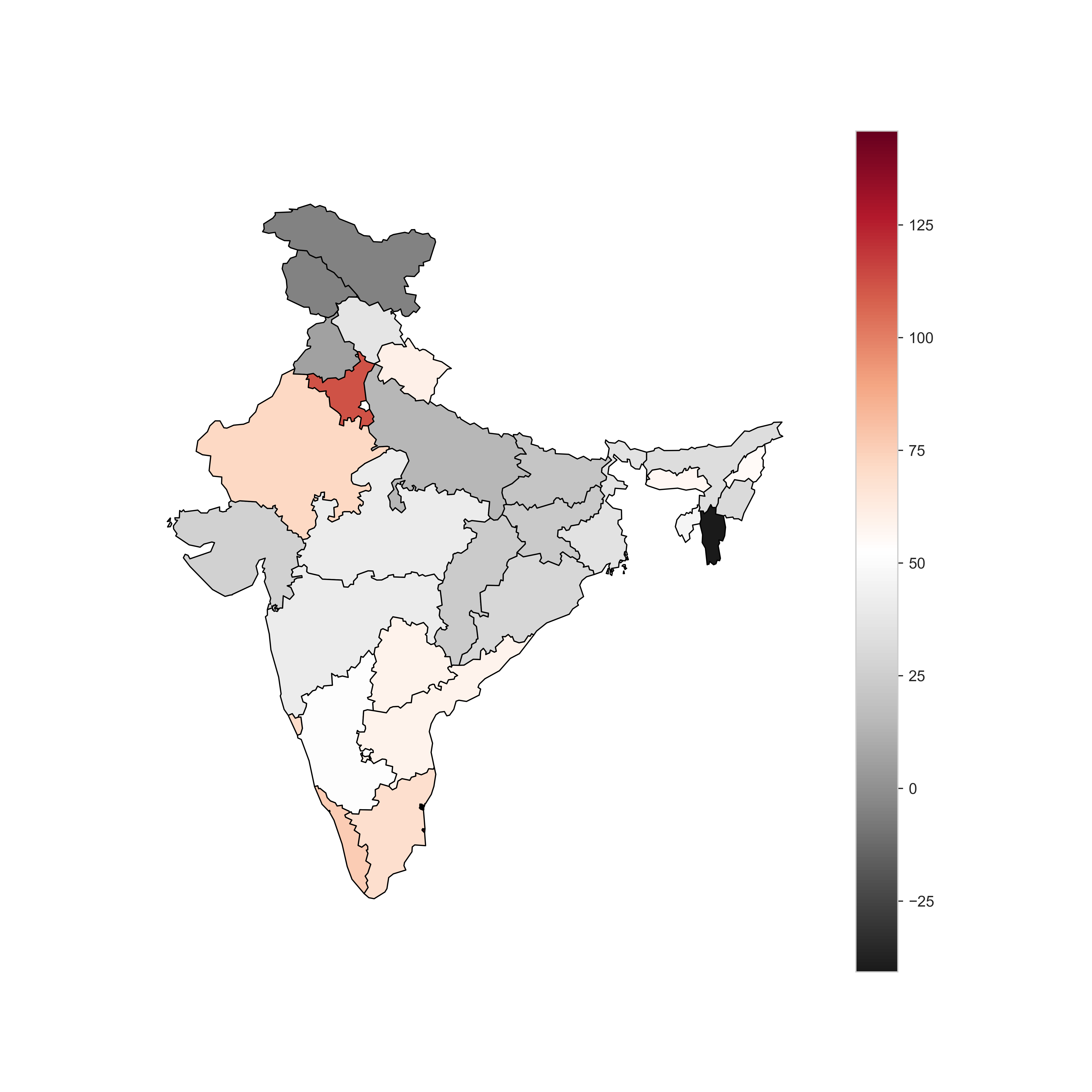}
        \caption{Average state-wise changes in women's casual agricultural labor wage rates (2012 Rs.) between 2004-5 and 2011-12. Averages are unweighted. Source: \cite{IHDSI}}
        \label{subfig:aglabw_states}
\end{figure}
\begin{figure}[]
    \centering
        \includegraphics[width=0.7\linewidth]{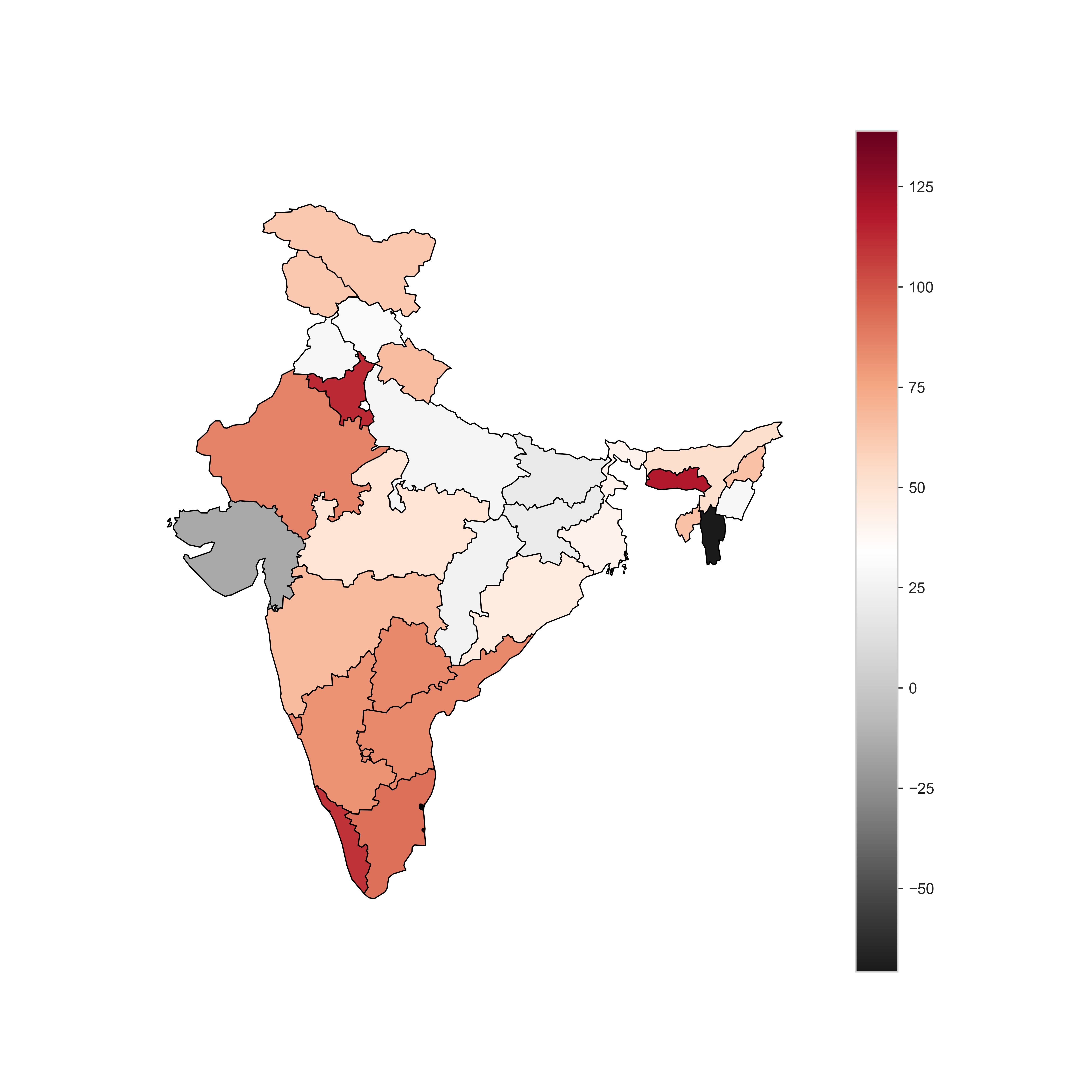}
        \caption{Average state-wise changes in men's casual agricultural labor wage rates (2012 Rs.) between 2004-5 and 2011-12. Averages are unweighted. Source: \cite{IHDSI}}
        \label{subfig:aglabm_states}
\end{figure}
\begin{figure}[]
    \centering
        \includegraphics[width=0.7\linewidth]{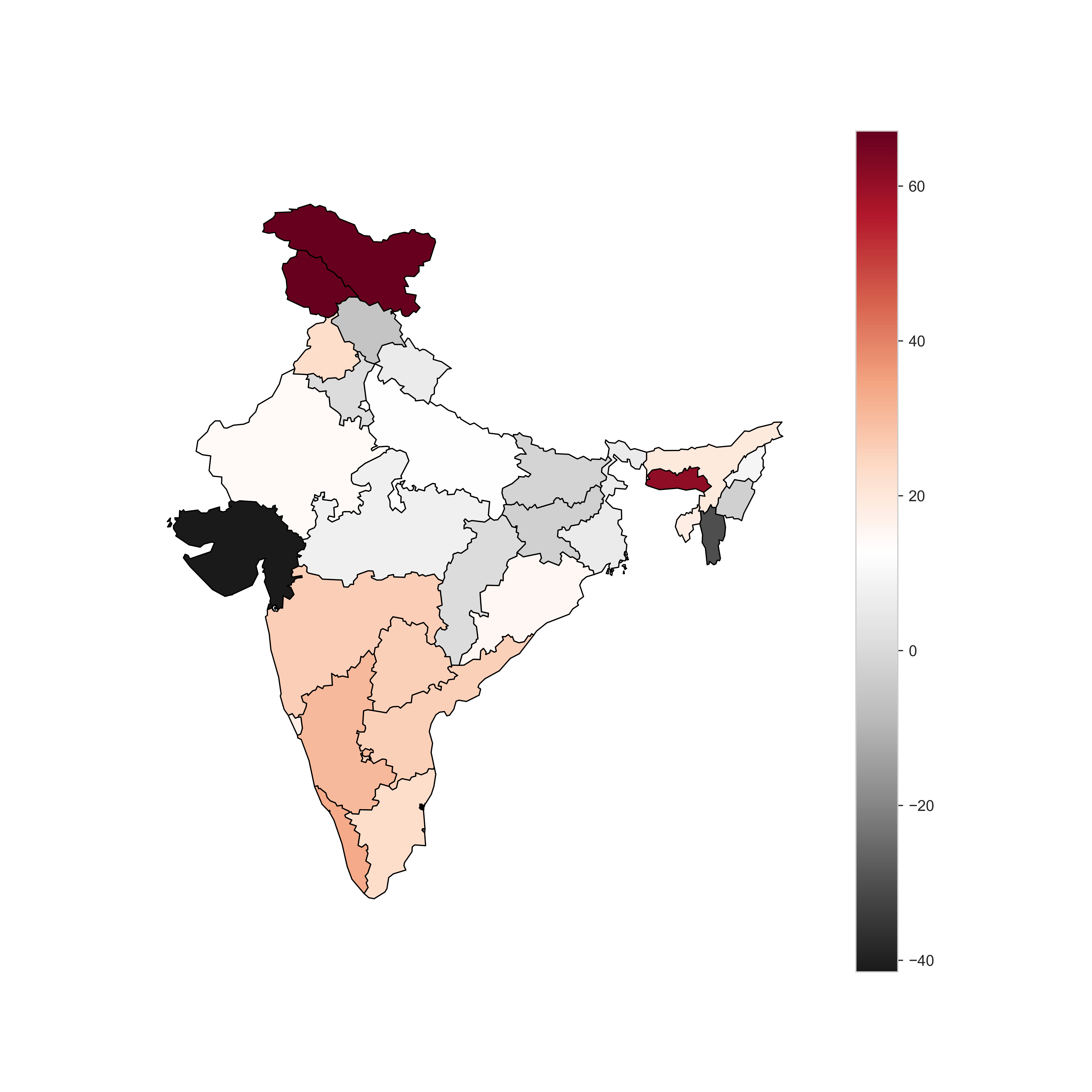}
        \caption{Average state-wise changes in the gender wage gap (men's rate - women's rate) (2012 Rs.) between 2004-5 and 2011-12. Averages are unweighted. Source: \cite{IHDSI}}
        \label{subfig:wagegap_states}
\end{figure}
\begin{figure}[]
    \centering
        \includegraphics[width=0.7\linewidth]{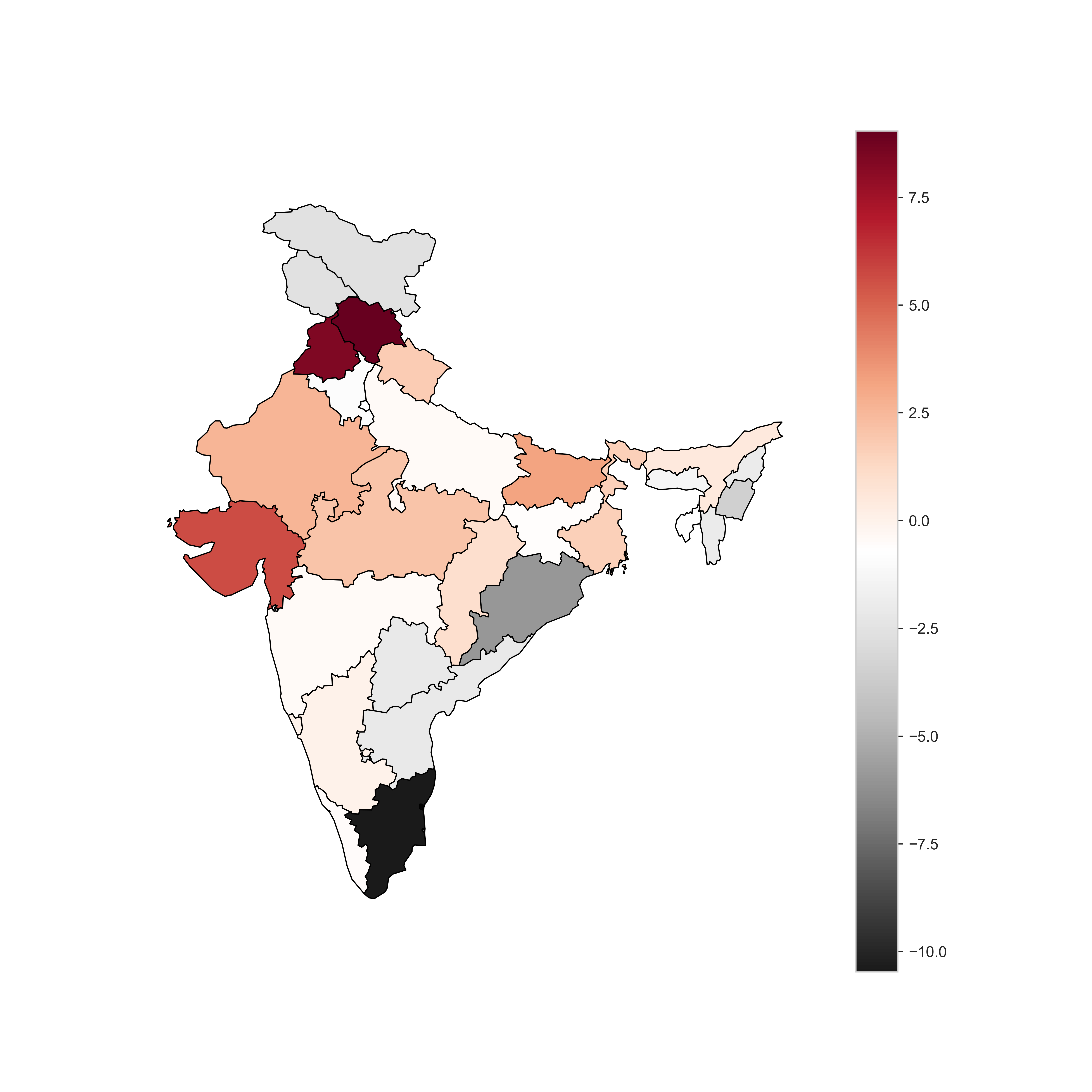}
        \caption{Average state-wise changes in reliability (hours per day) between 2004-5 and 2011-12. Averages are unweighted. Source: \cite{IHDSI}}
        \label{subfig:rel_states}
\end{figure}
Another method through which one could eliminate the possibility of selection bias in assignment to treatment due to covariates is by measuring the correlations between pre-existing levels of village characteristics and the treatment dose. Table \ref{tab:hh_rel} presents the mean and standard deviations of village characteristics of villages that receive more reliable electricity access and those that do not along with Pearson's correlation coefficient of the change in reliability with the 2005 levels of the variables. Overall, there are only marginal differences between most characteristics of treatment and control group villages, and the difference remains similar over time which encourages a difference-in-differences approach \citep{wing2018designing}. 


Since the correlations between pre-treatment characteristics and the change in reliability are weak, I assume that the treatment is random. This is not an unreasonable assumption as changes in reliability, are hardly conscious decisions made by policymakers based on village characteristics, and are mostly a result of transmission and distribution losses. A more formal analysis of whether the assignment to treatment is random or not is presented in Appendix \ref{appendix:random}.

In 2004-5, the average hours of electricity available to villages was 11.76 hours with a standard deviation of 7.80 hours, which increased to 13.38 hours in 2011-12, with a standard deviation of 6.74 hours. The correlation between household-level reliability and village-level reliability is 0.5491 and 0.6628, in 2004-05 and 2011-12, respectively, both coefficients being exceptionally statistically significant. While there is some parity between household responses on reliability, there are large variations as well, with households often even reporting higher quality than that received at the village level, which may indicate that the data at the household level for electricity quality may not be as trustworthy.  

\begin{table}[t]
\color{black}
    \centering
   \resizebox{\textwidth}{!}{ \begin{tabular}{cccccccccc}
    \hline 
     &&&&&&&&&\\
         && \multicolumn{8}{c}{Means (standard deviation)  }   \\ 
          &&&&&&&&&\\
          \cline{3-10}
      &&&&&&&&&\\
         && \multicolumn{2}{c}{2004-2005} && \multicolumn{2}{c}{2011-2012}&& \multicolumn{2}{c}{$\Delta$}\\
           &&&&&&&&&\\
         \cline{3-4} \cline{6-7} \cline{9-10}
 &&&&&&&&&\\
         && Increase in & Non-increase in && Increase in & Non-increase in &&Increase in & Non-increase in \\ 
         && Reliability & Reliability && Reliability & Reliability && Reliability & Reliability \\
          &&&&&&&&&\\
          \hline 
           &&&&&&&&&\\
         
        &&&&&&&&&\\
          \multirow{1}{*}{Women's daily wage rate (2012 Rs.)} & summer-monsoon  & 86.73  & 85.94  && 134.2  &  138.9  && 41.23 &46.53 \\
         && (39.8) & (42.51) && (59.26) & (70.89)&& (48.73)   & (50.46)\\
         & Winter & 84.56  & 85.63  && 132.93  &  138.86  && 40.02  & 48.01 \\
         && (40.91) & (42.399) && (56.43) & (68.46)&&  (48.19)   &(52.45)\\
         &&&&&&&&&\\
         \multirow{1}{*}{Men's daily wage rate (2012 Rs.) }& summer-monsoon & 116.99   & 119.66  && 168.26  &  183.48 && 51.24  &61.93 \\
         && (51.51) & (52.81) && (68.21) & (83.76)&& (57.66)   & (63.38)\\
         & Winter & 115.95  & 117.95  && 166.76  &  184.65  && 49.48  & 63.51 \\
         && (53.04) & (51.8) && (66.04) & (83.15)&& (56.72)    &(61.76)\\
         &&&&&&&&&\\
         
       \multirow{1}{*}{Average daily wage rate (2012 Rs.)} & summer-monsoon & 105.37  & 105.46  && 151.52  &  161.73  && 45.53 &54.75 \\
         && (46.86) & (46.57) && (61.01) & (74.74)&&  (51.71)  & (56.00)\\
         & Winter & 105.23  &104.74 && 150.13  &  162.96  && 42.8  & 55.58 \\
         && (49.01) & (46.66) && (58.52) & (73.85)&& (51.1)    &(54.81)\\
         &&&&&&&&&\\
        &&&&&&&&&\\
         \hline 
    \end{tabular}}
    \caption{ \color{black}Casual agricultural labor wage rate statistics for villages where reliability improves and villages where it does not, segregated by gender and cropping seasons. Reliability is measured as the average number of hours of electricity available per day. Includes 1281 Indian villages. The wages have been standardized to the 2012 Rupee by multiplying the 2005 rates by the national consumer price index (reported by the World Bank). Averages are unweighted. The data are unweighted. Source: IHDS I, and IHDS II surveys. \color{black}}
    \label{tab:wages_stats_season}
    \color{black}
\end{table}

The IHDS surveys have data on casual agriculture wage rates in villages for both major cropping seasons in India - the Kharif season where crops are sowed towards the end of the summer, around June, and harvested at the end of the monsoon around October and November, and the Rabi season in which crops are sown at the start of winter, around mid-late November, and harvested in Spring, in April. Rice and maize are the main Kharif crops in India, while wheat is the main Rabi crop. I construct the wage rates for each category by averaging the wage rates for sowing and harvesting, when available\footnote{In the absence of one of the two, I use the wage rate for the task that is available. Plowing wage rates were also present in the first round of the survey but I excluded plowing wage rates from the outcome variable, as plowing was relegated to the category of ``Other agricultural work", a category which may include wages for other work such as winnowing or threshing.}. From the data (Table \ref{tab:wages_stats_season}), I can observe that there is little seasonal variation in the wage rates. However, the wage rates are usually higher for men than for women, with men earning nearly 35\% more than women in 2004-5. Even though the wage gap shrinks slightly by 2011-12, there is still a considerable disparity. Figure \ref{subfig:aglabw_states}-\ref{subfig:rel_states} shows the distribution of changes in real wage rates for agricultural labor for women (Figure \ref{subfig:aglabw_states}), men (Figure \ref{subfig:aglabm_states}), the gender wage gap (Figure \ref{subfig:wagegap_states}), and reliability (Figure \ref{subfig:rel_states}). Typically, in states where reliability reduces, the gender wage gap increases, while in states where reliability increases, the gender wage gap reduces (Rajasthan, Punjab, and Northeast Indian states being the only examples), with the correlation between changes in state-wise reliability and the gender wage gap being -0.33. It is noteworthy that the reliability-associated differences in the growth of wages discussed earlier in Section \ref{sec:introduction}
exist in seasonal wages as well. The difference, however, is less for women's wage rates, than men's wage rates, suggesting that gender may play a role in the dynamics of the effect on wage rates. 
\section{Empirical Strategy} \label{sec:Methods}

\subsection{Difference-in-differences specification}
I begin with a two-way fixed effects specification for village $j$: 
\begin{equation}
\label{eq:levels}
    y_{jt} = \alpha_j + \beta R_{jt} + \gamma X_{jt} + \mu_t   + \epsilon_{jt}, ~~~ t=2005, 2012
\end{equation}
where the outcome variable $y_{jt}$ can be either the log or level of the casual agricultural labor wage rate of the village, $R_{jt}$ is the electricity reliability, $X_{jt}$ is the vector of control variables, $\alpha_j$ is the village-specific fixed effect, $\mu_t$ is the time-specific fixed effect, and $\epsilon_{jt}$ is the idiosyncratic error term. The control variables that I use are the fraction of households in the village with electricity access, whether there are metalled roads, NGOs or development organizations, primary healthcare centers in the village, the distances to the closest bank or credit cooperative, and market, the number of government and private primary, middle, secondary and higher secondary schools, and whether the village has less than 1000 inhabitants or more than 5000 inhabitants. I am interested in the coefficient $\beta$ which quantifies the effect of reliability. Taking the first difference of equation \ref{eq:levels} in time, 
\begin{equation}\label{eq:diff_primitive}
    \Delta y_{jt} = \Delta \mu_t + \beta \Delta R_{jt} + \gamma \Delta X_{jt}    + \Delta \epsilon_{jt}, ~~~ t=2012
\end{equation}
 Since I have data on wage rates, reliability, and control variables, that allow me to generate a cross-section of differences as required for the above model, I can estimate the coefficients using an ordinary least squares (OLS) regression, with $\Delta \mu$ as the intercept. 

At this stage, it is important to note that the reliability is constrained to be between 0 and 24 hours a day. Therefore even if the assignment is random, only villages with access to high-quality electricity in the pre-treatment period would be able to experience a large negative treatment, and similarly, only villages with poor reliability in the pre-treatment period will be able to experience large benefits. 
This association poses a problem to the identification, as it will be difficult to determine if the effects are attributed to the changes or rather a lagged effect arising from pre-treatment levels. Is an increase in reliability reducing wages, or are wages reducing because the village historically has low reliability? One simple way to deal with this problem is to include the pre-treatment level of reliability in the regression as well, so that the regression does not suffer from omitted variable bias, and the effects can be independently attributed to the change in reliability and the pre-treatment level of reliability. Incorporating a lagged effect of the pre-treatment level ($R_{jt-1}$), the model stands as follows: 
\begin{equation}\label{eq:diff_in_diff}
     \Delta y_{jt} = \Delta \mu_t + \beta_1 \Delta R_{jt} + \beta_2 R_{jt-1} + \gamma \Delta X_{jt}    + \Delta \epsilon_{jt}, ~~~ t=2012
\end{equation}
Including the pre-treatment level in the difference equation, is the same as including a partial sum of reliability in the original two-way fixed effects specification. Thus, the correct form of equation \ref{eq:levels} is 
\begin{equation}
       y_{jt} = \alpha_j + \beta_1 R_{jt} + \beta_2 \sum_{s=t_0}^{t-1} R_{js} + \gamma X_{jt} + \mu_t   + \epsilon_{jt}, ~~~ t=2005, 2012
\end{equation}
Clearly, the fraction of households connected, which is a control variable, has the same problem, as values range from 0 to 1, and thus I also incorporate a pre-treatment level of the fraction as a control variable. The coefficient $\beta_1$ in Equation \ref{eq:diff_in_diff}, would give an unbiased estimate of the effect of changes in reliability.

However, there is little reason to believe that the effects of reliability would be symmetrical, i.e., the effects of a positive change in reliability need not quantitatively be the opposite of the effects of a negative change in reliability. Therefore, I consider two separate natural experiments, one to study the effect of improvements in reliability, and one to study the effect of a decline in reliability. The advantage of splitting the sample into villages that include only positive or negative samples is that I require changes in reliability -- the treatment -- to be randomly assigned only within each sample. For instance,  there could be a selection bias in determining which villages receive positive and negative treatment, but as long as the magnitude of the treatment is random, the results will hold. In Appendix \ref{appendix:random}, I investigate whether the treatment is genuinely random. We find that for villages where reliability increases, the treatment can be assumed to be as good as random, and the results of the difference-in-differences estimates will be unbiased. However, for villages where reliability reduces, there may be some endogeneity in the assignment to treatment and the results may not be as robust. Therefore, when we split the sample, as opposed to including two treatment variables in a single regression, we are able to isolate a sample of villages and a type of treatment that is as good as random.

\subsection{Mechanisms: Labor Demand and Supply}

Since there are no village-level variables for demand and supply, I use household-level data to study the demand and supply of labor, although I use village-level reliability as the treatment, as households often report considerably different levels of electricity reliability and the household data may not be as accurate.\footnote{Since all the electricity comes from a common distributor in the village, the variation in quality is unlikely to be as much as I find in the household data. While there may be losses that affect individual households, sometimes households report levels of electricity reliability that exceed the levels received by the village, which indicates that household reporting may be an unreliable source of information for electricity quality. Furthermore, even load shedding is typically carried out for entire villages and not individual households.} Moreover, the quality of electricity available to houses may not be the same as the quality of electricity available to farms (which will be useful in studying labor demand), and in the absence of data on electricity reliability in farms, the village's reliability is the best proxy. 

There is no variable in the survey that can be directly used as a measure of labor supply. Thus, I investigate a channel through which labor supply may be affected, as this is easier to test from the data. I postulate that the supply of labor would be affected by electricity quality primarily through a reduced time burden of domestic chores. In particular, the tasks that would be especially affected by electricity disruptions are lighting and refrigeration. 4045 households in the sample (21.2\%) claimed to own refrigerators in 2011-12. Long disruptions would hamper refrigeration, and the presence of good-quality electricity could reduce the time and fuel required for cooking, through refrigeration. Since the survey does not have questions related to the time spent cooking, I am constrained to restrict my analysis to fuel. About 75\% households use firewood for either cooking or lighting. Over 40\% households also use dung for cooking. A large number of households collect these fuels from either owned or community land, good-quality electricity could significantly reduce the time spent by members in collecting biomass. Thus, as a supply channel, I study the effect of electricity reliability on the time spent by men and women on fuel collection. I use a difference-in-difference specification, similar to equation \ref{eq:diff_in_diff}. For a household $i$ in village $j$, we model the time spent in fuel collection $F_{ijt}$ by 
\begin{equation}\label{eq:fuel collection}
    \Delta F_{ijt} = \Delta \mu_t + \beta_1 \Delta R_{jt} + \beta_2 R_{jt-1} + \gamma \Delta X_{ijt}    + \Delta \epsilon_{ijt}, ~~~ t=2012
\end{equation}
where most of the coefficients represent the effects of the same quantities as before (on time spent in fuel collection), except $\gamma'$ which now includes coefficients for some household-level controls along with the village-level controls listed about. The household-level controls are the number of adult men and women, the presence of a source of drinking water in the house,  the presence of flush toilets, and the presence of separate kitchens. Obviously, village-level electricity quality cannot have a direct effect on households not connected to the grid. Hence, I only consider those households that had electricity in both 2004-5 and 2011-12. Ideally, had there been data for how much labor individuals in a household were willing to supply I could have studied the effect of time saved in fuel collection on such a variable. But in the absence of such a variable, a proper analysis of labor supply is beyond the scope of this work. 

Unlike supply, demand can be studied by considering the person-days of labor hired by agricultural households to work on their farms. Of course, I need to control for the labor wage rates as well, to avoid an omitted variable bias. Not controlling for labor wage rates could be particularly problematic, as reliability is correlated with the wage rates, and we know from theory that wage rates themselves can affect the demand for labor. In addition to the control variables listed above, we also include the change in the ownership of electric pumps ($\Delta P_{ijt}$). We use the logarithm of the person-days of hired farm labor $D_{ijt}$ as the outcome variable and formulate another difference-in-differences equation 
\begin{equation}\label{eq:demand}
    \Delta D_{ijt} = \Delta \mu_t + \beta_1 \Delta R_{jt} + \beta_2 R_{jt-1}    + \delta_m M_{jt} + \delta_w W_{jt}+ \phi \Delta P_{ijt} + \gamma \Delta X_{ijt} + \Delta \epsilon_{ijt}, ~~~ t=2012
\end{equation}
$M_{jt}$ is the log men's casual agricultural wage rate in village $j$, and $W_{jt}$ is the corresponding quantity for women's wages. The coefficients $\delta_m$ and $\delta_w$ are the elasticities of labor demand for men and women, respectively, and tell us the fractional change in demand, for a fractional change in wages. Since pump ownership could have an impact on demand, it is also important to check whether reliability has an effect on pump ownership and is not affecting demand through pump ownership. We do so using another difference-in-difference regression: 
\begin{equation}\label{eq:pump}
    \Delta P_{ijt} = \Delta \mu_t + \beta_1 \Delta R_{jt} + \beta_2 R_{jt-1}     + \gamma \Delta X_{ijt} + \Delta \epsilon_{ijt}, ~~~ t=2012
\end{equation}

\section{Results} \label{sec:Results}

\subsection{Effect of electricity quality on wage rates}
Table \ref{tab:results_aglab} presents the results of the difference-in-differences regressions for the impact of a change in the electricity reliability (both positive and negative) on the change in men's and women's casual agricultural labor wage rates. Although we present the results for both positive and negative treatments, only the positive treatments can be assumed to be as good as random, and the estimates of the effects of negative treatment may not be as robust. First, I find that reliability has a negative effect on both logs and levels of casual agricultural labor wages, for both men and women, although the effect on women's wage rates is considerably smaller, and not statistically significant. For every additional hour's increase in electricity reliability in a village, men's casual agricultural wage rates reduce by nearly Rs. 2 or 1.34\% in relative terms. The effect on both levels and logs is statistically significant at the 1\% level. On the other hand, positive treatments have small and statistically insignificant impacts on women's wage rates. 

The effects of a decrease in quality, however, has consistently larger effects on both men's and women's wages. For every additional hour's reduction in electricity availability, men's wage rates rise by over Rs. 2 (significant at the 1\% level), although the effect on the log of wage rates is small (Rs. 0.6) and not significant. Similarly, women's wage rates increased by Rs. 1.2 (significant at the 5\% level), although the effect on log wages is not statistically significant. Nevertheless, this implies that the effects of changes in reliability are asymmetrical, particularly for women's wage rates, and wage rates are more sensitive to reductions in reliability than increases. But these estimates are not as robust as those for improvements in reliability as the assignment to treatment may be biased by initial characteristics.
\begin{table}[t]
\centering
    \resizebox{\textwidth}{!}{
    \begin{tabular}{ccccccccccccc}
    \hline \\
    &  & \multicolumn{11}{c}{Coefficients (standard errors)} \\ 
    \\\cline{3-13}
    \\
    && \multicolumn{5}{c}{Positive Treatment}  && \multicolumn{5}{c}{Negative Treatment}  
    \\\cline{3-7}\cline{9-13}
    \\
    && \multicolumn{2}{c}{Women's Wage Rates}  && \multicolumn{2}{c}{Men's Wage Rates} &&  \multicolumn{2}{c}{Women's Wage Rates}  && \multicolumn{2}{c}{Men's Wage Rates} \\&&&&&&&&&&&&\\
    && \multicolumn{2}{c}{(2012 Rs.)}  && \multicolumn{2}{c}{(2012 Rs.)} &&  \multicolumn{2}{c}{(2012 Rs.)}  && \multicolumn{2}{c}{(2012 Rs.)} \\
    \\
    && \multicolumn{2}{c}{$N=658$}  && \multicolumn{2}{c}{$N=720$ } &&  \multicolumn{2}{c}{$N=580$}  && \multicolumn{2}{c}{$N=637$ } \\
    \\ \cline{3-4} \cline{6-7} \cline{9-10} \cline{12-13}
    \\
    &Independent Variable & Levels & Logs && Levels & Logs && Levels & Logs && Levels & Logs \\ 
    \\\hline 
    \\
           
     & Intercept&29.2885$^{***}$    & -0.2267$^{***}$  &&51.7882$^{***}$     & -0.0721              && 13.2521            & -0.3864$^{***}$       &&  29.7162$^{***}$      &-0.2218$^{***}$    \\ 
    &&(6.8389)&(0.0642)&&    (8.1564)   &(0.0591)&& (9.2220)   & (0.0663)   && (9.5753)   &(0.0547)\\
    &$\Delta$ Reliability (hours) &-0.7196             &-0.0060           &&   -1.9521$^{***}$     & -0.0134$^{***}$ &&  -1.1836**           & -0.0065         && -2.0438$^{***}$       &-0.0061   \\
    &&(0.5177)&(0.0044) &&  (0.6213)       &(0.0042)&& (0.5991)   &  (0.0045)  && (0.7467)   &(0.0049)\\
    & Pre-treatment Reliability (hours) &-0.6470              &-0.0047          &&    -1.4216$^{***}$    &  -0.0104$^{***}$  && -1.1436$^{**}$    &  -0.0050         && -1.7940$^{***}$       &-0.0091$^{***}$    \\
    &&(0.3962)&(0.0032) &&   (0.5042)      &(0.0032)&&  (0.4914)  & (0.0033)   && (0.5267)   &(0.0029)\\
    \\
    \hline 
    \end{tabular}}
    
\caption{The impact of reliability on men's and women's casual agricultural labor wage rates. Positive treatment refers to the set of villages where reliability increased or stayed the same. Negative treatment refers to the set of villages where reliability decreased or stayed the same. Robust standard errors clustered at the district level. \\ \small{*Significant at the 10\% level, 
    **Significant at the 5\% level, 
    ***Significant at the 1\% level}}
    \label{tab:results_aglab}
\end{table}

Interestingly,  even the pre-treatment level of reliability has a statistically significant negative effect on men's wages-- a Rs. 1.4 or a 1.04\% reduction for every hour's increase or a Rs. 1.8 or 0.91\% increase for every hour's reduction (all significant at the 1\% level).  From this, I can conclude that the effect of the overall level, not the change alone, is negative. However, for women's wages, the effect is again statistically insignificant for all cases except on the levels of the wage rate in the negative treatment case, where wages increase by Rs. 1.4, significant at the 5\% level. An obvious consequence of different effects on men's and women's wages is that better reliability leads to a smaller wage gap. From the estimated coefficients, it is clear that the wage gap increases over time for both men and women, after controlling for various factors, although villages where reliability improves would experience a smaller widening of the gender wage gap when compared to a village where reliability declines.

The effect of reliability on wage rates, however, is difficult to explain intuitively and will require studying the effects of reliability on labor demand and supply. 

\subsection{Time spent in fuel collection: A potential labor supply channel}
In order to study the possible effect of electricity reliability on the supply of labor, I estimate the impact of changes in reliability on the time spent by men and women on fuel collection. Table \ref{tab:results_supply} presents the results.\footnote{We include only villages where reliability increases as the treatment in these villages can be assumed to be random. See Appendix \ref{appendix:random}.} Reliability has a negative effect on the time spent on fuel collection. In villages where reliability improved, every additional hour of power available reduces the time spent by women on fuel collection by 12.6 minutes per week, which is statistically significant at the 1\% level, while men's fuel collection time reduces by 6.1 minutes per week, significant at the 10\% level. This can amount to a very large reduction of fuel collection time if the quality changes sizeably. 

\begin{table}[h]
    \centering
      \resizebox{\textwidth}{!}{ \begin{tabular}{cccccc}
    \hline &&&&&\\
           & \multicolumn{5}{c}{Coefficients (standard errors)} \\ 
         &&&&&\\\cline{2-6}
&&&&&\\
& \multicolumn{1}{c}{$\Delta$ Women's Fuel Collection Time}  & \multicolumn{1}{c}{$\Delta$ Men's Fuel Collection Time} &&\\ 
& \multicolumn{1}{c}{(minutes per week)}  & \multicolumn{1}{c}{(minutes per week)} &&\\
         & \multicolumn{1}{c}{$N=2667$}  & \multicolumn{1}{c}{$N=2513$ } && \\
         &&&&&\\
         \hline 
         &&&&&\\
          Intercept &  63.9049       &  -91.7331*              && \\
          &  (79.3594)     &  (53.9066)        &&  \\
          $\Delta$ Reliability (hours)&  -12.6077***        &  -6.1298*           && \\
          &    (4.2664)   &  (3.4914)        &&\\
          Pre-treatment Reliability (hours) & 0.6603        &  -0.3436      && \\
          &  (3.0550)     &   (2.7551)       && \\
          &&&&&\\
         \hline 
        \end{tabular}}
    \caption{The impact of reliability on men's and women's fuel collection times. Includes villages where reliability did not reduce. Robust standard errors clustered at the district level. \\ \small{*Significant at the 10\% level, 
    **Significant at the 5\% level, 
    ***Significant at the 1\% level}}
    \label{tab:results_supply}
\end{table}

Clearly, women's fuel collection times are affected more, as fuel collections like other domestic chores are typically carried out by women. Time relieved from household chores could substantially increase the supply of labor and has been hypothesized to be a channel for electricity-facilitated increase in labor participation in past studies as well \citep{dinkelman2011effects}. However, it may be more complicated, especially in the Indian context, and in other overwhelmingly agrarian rural economies. In general, time saved could increase women's labor market participation, which would increase the supply of labor and may have a detrimental effect on wages, or the time could be utilized in leisure, in which case the labor market would be virtually unaffected.\footnote{Unfortunately, there is no variable that aptly quantifies leisure time, and I cannot test this empirically.} A third possibility is that women take up more domestic chores that used to be performed by men, freeing up men's time burden of domestic work, which could increase men's labor market participation, which would also increase labor supply. This may explain why men's wage rates fall more than women's wage rates. Although that could also be if women are hired more in the labor market because farmers may want to hire cheaper labor. But this possibility only works on the assumption that men's and women's labor is thought of to be fungible, which is unlikely to be the case. Another alternative is that women could provide labor on household-owned businesses, or more likely in the case of rural agricultural households in India, women could supply labor to farms owned by their households, a common phenomenon in Indian farms. While this would not exactly increase labor supply in a strict sense, since it is not paid work, it could reduce the demand for hired labor. Thus, it is also important that I study if reliability, or even fuel collection time, has an impact on labor demand. 

\subsection{Labor demand effects of electricity reliability}
To estimate the effects of electricity quality on labor demand, I study how changes in reliability affected the number of person-days of agricultural labor hired by farms. Table \ref{tab:results_demand} presents the estimated effects on the levels and logs of the number of days of hired work. Both the changes in reliability and the pre-treatment level of reliability have no statistically significant impacts on the number of days of labor hired by farms. We also do not find any statistically significant effects of changes in wage rates, either at the levels or logs, on demand. This suggests that the own elasticity of labor demand is inelastic, for both men and women. 
\begin{table}[h]
    \centering
      \resizebox{\textwidth}{!}{ \begin{tabular}{ccc}
    \hline &&\\
           & \multicolumn{2}{c}{Coefficients (standard errors)} \\ 
           &&\\\cline{2-3}
           &&\\
           & \multicolumn{2}{c}{$\Delta$ Number of Man-days of Hired Farm Labor} \\ 
         &&\\\cline{2-3}
&&\\
$N=593$& \multicolumn{1}{c}{Levels}  & \multicolumn{1}{c}{Logs} \\ 
         &&\\
         \hline 
         &&\\
          Intercept & -39.3914     &  -0.9263***      \\
          &  (26.9972)     &  (0.2527)          \\
          $\Delta$ Reliability (hours)&  -0.0521       &  0.0104                  \\
          &    (1.0093)   &  (0.0160)        \\
          Pre-treatment Reliability (hours) & 0.9098        &  0.0055       \\
          &  (0.8845)     &   (0.0127)        \\
          $\Delta $ Women's agricultural labor Wage Rate (2012 Rs.)& 0.2541*           &        \\
          &   (0.1374)      &              \\
          $\Delta $ Men's agricultural labor Wage Rate (2012 Rs.) & -0.1100          &        \\
          &  (0.1236)       &              \\
          $\Delta $ Log Women's agricultural labor Wage Rate (2012 Rs.)&    &      0.5450**         \\
          &         &      (0.2423)        \\
          $\Delta $ Log Men's agricultural labor Wage Rate (2012 Rs.) &    &        -0.4286        \\
          &         & (0.2850)             \\
          $\Delta $ Number of Electric Pumps Owned &56.9618***    &  0.3035**          \\
          &   (15.1355)      &     (0.0827)         \\
         \hline 
        \end{tabular}}
    \caption{The impact of reliability on the person-days of hired farm labor. Includes villages where reliability did not reduce. Robust standard errors clustered at the district level. \\ \small{*Significant at the 10\% level, 
    **Significant at the 5\% level, 
    ***Significant at the 1\% level}}
    \label{tab:results_demand}
\end{table}

\begin{table}[t]
    \centering
      \resizebox{\textwidth}{!}{ \begin{tabular}{ccc}
    \hline &&\\
           & \multicolumn{2}{c}{Coefficients (standard errors)} \\ 
           &&\\\cline{2-3}
           &&\\
$N=4410$           & \multicolumn{2}{c}{$\Delta$ Number of Electric Groundwater Pumps Owned} \\ 
         &&\\
         \hline 
         &&\\
          Intercept & \multicolumn{2}{c}{0.0985}      \\
          &  \multicolumn{2}{c}{(0.0655)}        \\
          $\Delta$ Reliability (hours)& \multicolumn{2}{c}{  -0.0050         }                  \\
          &    \multicolumn{2}{c}{(0.0038)      }      \\
          Pre-treatment Reliability (hours) & \multicolumn{2}{c}{ 0.0003         }     \\
          &  \multicolumn{2}{c}{  (0.0026)    }        \\
          &&\\
         \hline 
        \end{tabular}}
    \caption{The impact of reliability on the ownership of electric pumps. Includes villages where reliability did not reduce. Robust standard errors clustered at the district level. \\ \small{*Significant at the 10\% level, 
    **Significant at the 5\% level, 
    ***Significant at the 1\% level}}
    \label{tab:results_pump}
\end{table}

An increase in the number of pumps, however, has a large statistically significant positive impact on the demand for labor. Although the distribution of pumps may not be random thus making the sizes difficult to interpret, the positive effect of pump ownership on log changes in the number of hired days is significant at the 1\% level. Pump ownership increases the number of person-days of hired labor by 57 days or 30\% in relative terms, which is close to two months for one laborer. This effect does not disappear even after controlling for baseline levels of household affluence using the 2004-5 levels of expenditure. The positive effect, signifies, that electrical equipment such as irrigation pumps do not supplant labor, and instead drive up the demand for labor. This increase in demand could be because pumps may require labor to operate, or because pumps could increase agricultural productivity, incentivizing higher investments in agriculture. It could also arise from pumps facilitating agriculture during the dry seasons, which is most likely. Thus, the possibility that electricity quality reduces labor wages through the mechanization of agriculture may be ruled out. To ensure that reliability does not affect labor demand through increased pump ownership, I test the effect of changes on reliability on pump ownership. Table \ref{tab:results_pump} presents the results, and clearly there is no statistically significant increase (or decrease) in pump ownership due to increases in reliability.

\section{Conclusion}
I use village and household-level data from rural India to study how agricultural labor wages responded to improvements in electricity quality between 2004-5 and 2011-12. Using the India Human Development Survey which includes a panel of 1406 villages, I employ a difference-in-differences design to eliminate village and time-fixed effects. The major finding is that villages where electricity quality increases substantially experience a smaller increase in casual agricultural labor wage rates than villages where there are smaller or no improvements in electricity reliability. The effect is larger and statistically significant for men's wage rates, while the effect on women's wage rates is statistically insignificant, although still negative. As a result, villages with large improvements in electricity quality see a smaller widening of the gender wage gap, both in absolute and relative terms. This effect of electricity quality on wages also helps explain the negative effect of reliability on non-farm income found by \citet{samad2016benefits}.

Since changes in electricity reliability could affect wage rates through labor and supply, I also investigate such possibilities. In the absence of appropriate variables for supply, I study a channel that could potentially affect supply instead -- the effect of reliability on the time spent by men and women in fuel collection, hypothesizing that a smaller time burden of household chores could encourage greater participation in labor markets. I find that the effect of reliability on women's fuel collection times is statistically significant (and on men's times to a lesser extent). Women's fuel collection times are reduced more for every additional hour of electricity availability, as they already spend a greater time in fuel collection in comparison to their male counterparts. Despite the smaller time saved by men, the effect on men's wages may be larger because the time saved by women may not necessarily be spent on employment opportunities, and women may simply displace men in other domestic chores that are not affected by better-quality electricity. An unrelated implication of negative effects on the time spent in collecting primitive fuels is that it suggests that better quality electricity may facilitate households' assent on the fuel ladder \citep{vandrekroon2013}.

Since a smaller time burden need not necessarily initiate wage work, and households may also work in household farms, thereby driving down demand, it is also important to investigate how the demand for hired agricultural labor responds to improvements in electricity quality. I find no effect of electricity reliability on the number of person-days of hired work, implying that while supply increases, the demand does not change with electricity quality. Since demand does not increase, surplus labor aggravates the levels of disguised unemployment and drives down wages. Thus, better quality electricity indirectly causes an overall negative effect on the agricultural labor market and labor wage rates. 

What does, however, increase the demand for labor is the adoption of electric groundwater pumps, as these would require more hired labor to operate, and can lead to an increase in the number of cropping seasons driving up labor demand during the off-season. This has serious policy implications for governments in the Global South. Interventions such as household electricity connections or improvements in the quality of electricity supplied that could increase labor participation may be detrimental to wage rates due to the already saturated labor markets with persistent disguised unemployment,  but the situation may be alleviated if improvements in electricity quality are also accompanied by other changes that could help absorb the surplus labor. Since the labor supply effects of electricity access and quality in rural areas are well-known, governments should focus on simultaneously enabling farms to make good use of that electricity through pumps and other infrastructure or invest in alternate avenues to proportionately increase the demand for labor.

\singlespacing
\setlength\bibsep{0pt}
\bibliography{bibliography.bib}

\appendix

\section{Testing the Randomness of Treatment Assumption} \label{appendix:random}
That the treatment is as good as random, i.e., there are no factors observable or otherwise that bias the assignment to treatment is a key identifying assumption for difference-in-differences analyses. However, this needs to be tested formally. The method typically used to control for selection bias due to covariates is propensity scores.  Propensity scores measure the probability of a unit receiving the treatment that it does based on covariates prior to the assignment of treatment. In this case, it quantifies the probability of a village experiencing the improvement or decline in the quality of electricity that it did. If there are no statistically significant coefficients associated with covariates, it means that the effect of covariates in biasing treatment assignment is indistinguishable from zero. If coefficients are statistically significant, it implies that the treatment may not be random. 

Clearly, the change in reliability, my treatment variable is a continuous variable potentially ranging from -24 hours to +24 hours. Therefore, I estimate the effect of covariates on the assignment of the treatment dose, which would be a precursor to estimating propensity scores if the treatment were non-random. Following \citet{hirano2004propensity}, I assume a normal distribution of the treatment ($T_i$) given the covariates ($X_i$):
\begin{equation}
    \label{eq:GPS}
    T_i|X_i=\mathcal{N}(\beta_0+\beta'X_i,\sigma^2)
\end{equation}
Here, $\beta_0$ and $\beta'$ can be found by estimating from the linear model 
\begin{equation}
    T_i = \beta_0 + \beta'X_i + \varepsilon_i
\end{equation}
and $\sigma$ is the standard deviation of the residuals $\varepsilon_i$. The control variables used in the vector $X_i$ are the fractions of various caste populations in the villages, the size of the villages, and the pre-treatment levels of the presence of metalled roads, the number of government schools, the distance from the nearest bank or credit organization, the distance from the nearest market, the presence of NGOs or development organizations, the presence of primary healthcare centers, the fraction of households connected to the grid, and the years since the village was first connected to the grid.

Table \ref{tab:RC_first_stage} presents the results of the first-stage regression used to estimate propensity scores. The four columns show the estimates for four different samples of villages for two different types of treatment -- the first is the sample of villages that had data for women's wage rates for villages where reliability improved and the second is the sample of villages where reliability improved that had data for men's wage rates. The third and fourth are the corresponding estimates for villages where reliability depreciated. The four separate estimates are necessary also because we estimate these effects separately in our main study. 

\begin{table}[t]
    \centering
    \resizebox{\textwidth}{!}{\begin{tabular}{c c c c c c}
    \hline 
&&&&&\\
       & \multicolumn{5}{c}{$\Delta$ Electricity Reliability} \\
&&&&&\\ \cline{2-6}
    
&&&&&\\
       & \multicolumn{5}{c}{Coefficient (standard error)} \\
&&&&&\\ \cline{2-6}
&&&&&\\
&\multicolumn{2}{c}{Positive Treatment}&&\multicolumn{2}{c}{Negative Treatment}\\
&&&&&\\\cline{2-3}\cline{5-6}
&&&&&\\
 &Sample of Women's Wage Rates & Sample of Men's Wage Rates&&Sample of Women's Wage Rates & Sample of Men's Wage Rates\\
&&&&&\\
Covariates & $N=643$& $N=705$&&   $N=579$   &   $N=636$   \\
&&&&&\\
         \hline 
&&&&&\\
    Intercept &   4.8532*      &   3.7443        &&     1.3095     &  0.2580     \\
    &    (2.7732)       &   (2.7822)        && (2.2233)&  (1.4103)  \\
    Fraction of population -- Brahmin (\%) & -0.0184     &   -0.0053        && -0.0356     &  -0.0326      \\
    &   (0.0310)        &  (0.0317)        && (0.0263)             &  (0.0221)  \\
    Fraction of population -- non-Brahmin forward castes (\%) & -0.0032      &   0.0057        &&-0.0201    &   -0.0092        \\
    &  (0.0284)         &    (0.0284)      &&   (0.0228)            &  (0.0155)  \\
    Fraction of population -- Other Backward Castes (\%) & -0.0156  & -0.0135         &&-0.0442**     &  -0.0336**      \\
    &    (0.0272)       &  (0.0270)          && (0.0219)& (0.0148)   \\
    Fraction of population -- Scheduled Castes (\%) &0.0038     &0.0229     &&-0.0378     &  -0.0179      \\
    &  (0.0283)         &   (0.0289)  && (0.0262)& (0.0180)   \\
    Fraction of population -- Scheduled Tribes (\%) &  0.0166     &0.0238     && -0.0401*    &  -0.0309*      \\
    &   (0.0294)        &  (0.0294)        &&(0.0239)& (0.0175)   \\
    Presence of metalled road & -0.2162     & 0.0685      &&0.0244     &  0.1114      \\
    &   (0.4076)        &   (0.4263)       && (0.4903)&    (0.4470)\\
    Number of government schools &-0.0673      &  -0.0790        &&-0.1347    &   -0.1600      \\
    &  (0.0836)         &  (0.0876)        &&(0.0950)&   (0.0934) \\
    Village population -- small& 0.0114     & -0.1607       &&-0.2980     &  -0.3623      \\
    &   (0.5696)        &   (0.5654)       &&(0.5220)& (0.4761)   \\
    Village population -- big& -0.8590      &  -0.6042      &&-0.4257    & -0.2428       \\
    &    (0.5351)       &   (0.5370)       &&(0.7113)&  (0.6733)  \\
    Distance from bank/credit organization&   0.0558     &   0.0549      && -0.0868    &  -0.0501      \\
    &     (0.0369)      &  (0.0366)        &&(0.0651)& (0.0535)   \\
    Distance from market(km) &  0.0141       &  -0.0070       &&-0.1143**   & -0.1052**       \\
    &  (0.0449)         & (0.0450)         &&(0.0456)&  (0.0433)  \\
    Presence of NGO/development organization& 0.5932     &  0.1416      &&-0.2576     &  -0.3167      \\
    &   (0.6640)        &  (0.6778)        &&(0.5496)&  (0.5424)  \\
    Fraction of households connected (\%)& -0.0114      &  -0.0047     &&0.0102     &  0.0091      \\
    &  (0.0071)         &  (0.0072)        &&(0.0092)&  (0.0083)  \\
    Years since the village was first connected&    0.0302*        &  0.0322**      &&-0.0523***   &   -0.0565***     \\
    &      (0.0164)        &  (0.0160)        &&(0.0169)& (0.0162)   \\
    Presence of primary healthcare center&0.7551     &   0.5502      &&-0.3163     &  -0.3212      \\
    &  (0.6095)         &   (0.6165)       &&(0.5049)&   (0.4725) \\
&&&&&\\
    \hline
    
    \end{tabular}}
    \caption{First stage results of the effect of 2004-5 levels of covariates on the treatment dose (change in reliability). Robust standard errors clustered at the district level.\\ \small{*Significant at the 10\% level, 
    **Significant at the 5\% level, 
    ***Significant at the 1\% level}}
    \label{tab:RC_first_stage}
\end{table}


From the table, we can see that in the villages that received a positive treatment, no variable is statistically significant at the 5\% level for the sample of villages where data on women's wage rate is available. For the sample where data on men's wage rates are available, only one variable is statistically significant -- the number of years since a village was connected -- that too being a borderline case ($p=0.0446)$. Since most of the villages are common between these two samples, this effect may be due to correlations arising from certain outliers. Furthermore, the effect is extremely small in size - less than 2 minutes for every year that the village has been connected for. To put this in context, one standard deviation in the years of connection (about 15 years) makes a difference of only 27 minutes, which is about a tenth of one standard deviation of changes in reliability in the sample. This is a very small effect, and not even consistently significant in the two samples receiving positive treatment. Therefore, it is fair to assume that the positive changes in reliability (and controls) were as good as random. It is noteworthy that contrary to expectations, I find that variables known to be important in terms of connecting villages such as size \citep{burlig2016out}, or caste compositions \citep{aklin2021inequality}, do not affect changes in the quality of electricity access in these villages. Interestingly, even the fraction of households already connected to the grid has virtually no effect on reliability.  

On the other hand, negative changes in reliability may not random. Multiple variables -- the fraction of other backward caste members, distance from markets, and the years since connection -- are all statistically significant at the 5\% level or higher. This indicates that there may be some selection bias due to covariates in determining how large the fall in quality is in these samples, although the sizes of these effects are small and in most cases may not make a large difference. Nevertheless, my results on negative treatments are bound to be less robust, than the results on positive treatment. Due to this asymmetry in the presence of biasing factors between positive and negative treatment assignment, I slice the samples into two -- one each for the two types of treatment, so that at least estimates from two out of four samples are robust.\footnote{I also attempted a propensity-score weighted regression for this set but the common support assumption may not hold and some households receive an extra-ordinarily high weights, in the order of several hundreds while an overwhelming majority of weights are much lesser.}

\end{document}